\documentclass[nofootinbib,manuscript]{revtex4}
\usepackage{amsmath}

\begin{document}
\title{Squeezing and temperature measurement in Bose-Einstein
  Condensates} 
\author{J. Rogel-Salazar} 
\affiliation{Quantum
  Optics \& Laser Science, Department of Physics, Imperial College,
  London SW7 2BW, U.K.}  
\author{S. Choi} \affiliation{Clarendon
  Laboratory, Department of Physics, University of Oxford, Parks Road,
  Oxford OX1 3PU, U.K.}  \author{G.H.C. New} \affiliation{Quantum
  Optics \& Laser Science, Department of Physics, Imperial College,
  London SW7 2BW, U.K.}  \author{K. Burnett} \affiliation{Clarendon
  Laboratory, Department of Physics, University of Oxford, Parks Road,
  Oxford OX1 3PU, U.K.}

\begin{abstract}
  \vspace{0.3in}In this paper we discuss the presence of
  temperature-dependent squeezing in the collective excitations of
  trapped Bose-Einstein condensates, based on a recent theory of
  quasiparticle damping. A new scheme to measure temperature
  below the critical temperature $(T\ll T_c)$ is also considered.
\end{abstract}
\maketitle

\section{Introduction}
\label{sec:introduction}

Following the first experimental observation of Bose-Einstein condensation
in 1995 \cite{1stexp}, analogies between this phenomenon and
non-linear optics have been pointed out \cite{4wm,reinhardt}.
Recently, squeezing of the matter waves has been analysed
\cite{duan,sorensen,EPR}; by considering the role of the
nonlinear interactions between the atoms in generating entangled atomic
beams, namely the spin-exchanging collisions of spinor Bose-Einstein
condensates (BEC's).

Here we analyse squeezing in relation to the coupling of {\em
  collective excitations} in a Bose condensed gas in a trap. Most of
the observed effects of elementary excitations can be explained using
the standard Bogoliubov theory \cite{bogo}. However, it cannot account
for higher order processes such as the Beliaev damping of the
excitations observed recently for a scissors mode of a BEC in
$^{87}$Rb \cite{beldamp}. Beliaev damping is a mechanism in which a
quasiparticle with energy $E_2$ interacts with the condensate and
generates two quasiparticles with energy $E_1$. The process is
actually based on a four-wave process that is desguised as a three-wave
interaction analogous to degenerate parametric down-conversion. The
new quasiparticles, being produced in pairs, constitute a squeezed
twin phonon beam. Recent theoretical development including the
presence of interactions between the quasiparticles \cite{sam}
provides an explanation for such processes. Finding quantitative
estimates for the amount of squeezing and entanglement based on this
theory is one of the main tasks of this paper.

Considering a system containing 10000 $^{87}$Rb atoms in a spherical
trap, we calculate the degree of squeezing of the {\em quasiparticle}
excitations. A related experimental challenge that we also address is
the accurate determination of temperature in BEC's at $T \ll T_c$. The
temperature is currently estimated by fitting a thermal Gaussian
profile to the atomic cloud. We propose a different way to determine
temperature by observing the variation of the envelope of collective
oscillations.

The paper is structured as follows: In Section \ref{sec:hamiltonian},
the Hamiltonian in both particle and quasiparticle basis is discussed
in relation to squeezing. The Hamiltonian in the quasiparticle basis
accommodates processes such as the Landau and Beliaev damping. In
Section \ref{sec:pdc}, we obtain equations of motion for the
quasiparticle modes and give quantitative estimates for the degree of
squeezing and entanglement in the quasiparticle excitations.  The
expected ``damping rate'' at various temperatures is calculated. This
is then used as a temperature measurement calibration for BEC. In
Section \ref{sec:discussion}, we conclude.

\section{Many-body Hamiltonian}
\label{sec:hamiltonian}

Within the second quantised formalism, the many-body Hamiltonian for a
system of bosons with binary interactions can be written as,
\begin{equation}
  \label{eq:bareH}
  \hat H= \sum_{ij} H_{ij}^{sp} \hat a^{\dag}_i \hat a_j+\frac{1}{2}  \sum_{ijkm} \langle ij|\hat V|km\rangle \hat a_{i}^{\dag} \hat a_{j}^{\dag} \hat a_{k} \hat a_{m},
\end{equation}
where the matrix elements $H_{ij}^{sp}$ are given by
\begin{equation}
  \label{eq:Hsp}
  H_{ij}^{sp}=\int d^3 {\bf r} \psi_i^*({\bf r})\hat H^{sp}\psi_j({\bf r}).
\end{equation}

Here, $H^{sp}=-\frac{\hbar}{2m}\nabla^2+V_{trap}$, is the
single-particle Hamiltonian with a confining potential $V_{trap}$, and
the basis state wave functions are $\psi_i ({\bf r})$. The matrix
element $\langle ij|\hat V|km\rangle$ denotes the matrix element for a
bare interaction potential $\hat V({\bf r})$ between two atoms; in
this work, we shall consider the typical delta-function contact
interaction potential for $\hat V({\bf r})$.  The operators $\hat
a_{i}^{\dag}$ and $\hat a_{i}$ are the creation and annihilation
operators for mode $i$ that obey the Boson commutation relations
\begin{equation}
  [\hat a_i,\hat a_j^{\dag}]=\delta_{ij},\text{\ \ \ \ \ } [\hat a_i,\hat a_j]=[\hat a_i^{\dag},\hat a_j^{\dag}]=0.\nonumber
\end{equation}

The Hamiltonian (\ref{eq:bareH}) is written in a single-particle basis
where the operator $\hat a_i$ annihilates a particle from the state
with wave function $\psi_i({\bf r})$. The wave function $\psi_0({\bf
  r})$ describes the condensate, whilst the remaining functions form a
complete set orthogonal to the condensate.

\subsection{Particle Hamiltonian}

Let us first investigate the condensate by concentrating only on the
condensate mode. This one-mode approximation applies well for
temperatures $T \simeq 0$ which is indeed the case for many
experiments. We find the Hamiltonian is then simplified to
\begin{equation}
  \hat H= \epsilon_{00} \hat a_0^{\dag} \hat a_0 + \frac{1}{2}\langle 00|V|00\rangle \hat a_0^{\dag} \hat a_0^{\dag} \hat a_0 \hat a_0.
\end{equation}
Treating the deviations from the ideal condensate as perturbations, we
are able to expand $\hat a_0$ as
\begin{equation}
  \label{pert}
  \hat a_0=z+\hat c,
\end{equation}
where $z=\sqrt{N_0}$ represents the mean-field (condensate), and $\hat
c$ is the small quantum perturbation. $\hat c$ may take the meaning of
non-condensed atoms. We choose the phase so that $z$ is real. The
mean-field approximation, although not exact, has been very successful
in describing a large number of experimental results. Using the
expansion for $\hat a_0$ and splitting our Hamiltonian into a Gaussian
and a non-Gaussian parts, we get the Gaussian part of our Hamiltonian
as
\begin{equation}
  \hat H=\epsilon_0\left[z^2+z (\hat c+\hat c^{\dag}) + \hat c^{\dag} \hat c\right]+\frac{\langle00|V|00\rangle}{2} \left[z^4  + z^2 (\hat c^2+ \hat c^{\dag 2})+2 z^3 (\hat c + \hat c^{\dag})+4 z^2 \hat c^{\dag} \hat c\right].
\end{equation}

The non-Gaussian part may be treated using perturbation theory
\cite{dunn}. The resulting Heisenberg equations of motion for $\hat c$
and $\hat c^{\dag}$ are
\begin{equation}
  \label{eq:heisenberg}
  \left( \begin{array}{c}
    \dot{\hat c}(t)\\ \dot{\hat c^{\dag}}(t)
  \end{array} \right) = {\rm i} \left( 
  \begin{array}{rr}
    -{\rm a} & -{\rm b}\\ {\rm b} & {\rm a}
  \end{array} \right) \left(
  \begin{array}{c}
    \hat{c}(t) \\ \hat{c}^{\dag}(t)
  \end{array} \right) +{\rm i} \left(
  \begin{array}{r}
    -\kappa\\ \kappa
  \end{array} \right),
\end{equation}
where ${\rm a}=\frac{1}{\hbar}\left(\epsilon_0+2 \langle 00|V|00
  \rangle z^2 \right)$, ${\rm b}=\frac{1}{\hbar} \langle 00|V|00
\rangle z^2$, and $\kappa=\frac{1}{\hbar}\left(\epsilon_0 z+\langle
  00|V|00 \rangle z^3\right)$.  The solutions of these equations are
\begin{equation}
  \label{eq:solution}
  \left( \begin{array}{c}
    {\hat c(t)}\\ {\hat c^{\dag}(t)}
  \end{array} \right) = \left( 
  \begin{array}{cc}
    A & B\\ B^* & A^*
  \end{array} \right) \left(
  \begin{array}{c}
    \hat{c}(0) \\ \hat{c}^{\dag}(0)
  \end{array} \right) + \left(
  \begin{array}{c}
    D\\ D^*
  \end{array} \right),
\end{equation}
where $A=\cos \Upsilon t - {\rm i} \frac{{\rm a}}{\Upsilon} \sin
\Upsilon t$, $B=-{\rm i} \frac{{\rm b}}{\Upsilon} \sin \Upsilon t$,
$D=\frac{({\rm a} - {\rm b}) \kappa}{\Upsilon^2} \cos\Upsilon t- {\rm
  i} \frac{\kappa}{\Upsilon} \sin \Upsilon t + \frac{({\rm b}-{\rm a})
  \kappa}{\Upsilon^2}$, and $\Upsilon=\sqrt{{\rm a}^2-{\rm b}^2}$. It
is important to note that the above equations are formally equivalent
to dynamic equations that describe parametric down conversion in
quantum optics \cite{barnett} if $\kappa \equiv 0$.

Defining the quadrature operators as $\hat X_p(t)=\frac{\hat c (t)+
  \hat c^\dag(t)}{2}$ and $\hat Y_p(t)=\frac{\hat c (t) - \hat c^\dag
  (t)}{2{\rm i}}$, we have that the variances can be written as
\begin{eqnarray}
  (\Delta \hat X_p(t))^2 &=& \frac{(2 \tilde n +1)}{4}\left(\cos^2 \Upsilon t+\frac{({\rm a}-{\rm b})^2}{\Upsilon^2}\sin^2 \Upsilon t \right)\\
  (\Delta \hat Y_p(t))^2 &=& \frac{(2 \tilde n +1)}{4}\left(\cos^2 \Upsilon t+\frac{({\rm a}+{\rm b})^2}{\Upsilon^2}\sin^2 \Upsilon t\right)
\end{eqnarray}  
where $\tilde n$ is the number of excited particles. In this case, the
magnitude of ${\rm a}$ and ${\rm b}$, or equivalently the magnitude
and phase of the mean field, $z$, control the amount of squeezing of
the particule quadrature operators.

\subsection{Quasiparticle Hamiltonian}
On the other hand, it is possible to describe the system including the
higher order processes. The dominant ones correspond to the quadratic
Hamiltonian $\hat H_Q=\hat H_0+ \hat H_1+ \hat H_2$, where the
subscripts 0,1,2 indicate the order in $\hat a_i$. This may be
diagonalised exactly using the standard Bogoliubov transformations,
changing the Hamiltonian to a quasiparticle basis where the
quasiparticle operators $\hat \beta_i$ are defined by\footnote{For
  simplicity from now on we drop the explicit notation in the
  operators}

\begin{equation}
  \label{eq:quasi}
  \beta_i=\sum_{j \neq 0}U_{ij}^* a_j-V_{ij}^* a_j^{\dag}.
\end{equation}

Here $U_{ij}$ and $V_{ij}$ are the well-known matrices associated with
the Bogoliubov transformations, obeying the orthogonality and symmetry
conditions $UU^{\dag}-VV^{\dag}=1$ and $UV^T-VU^T=0$. These matrices
have to be evaluated numerically depending on temperature and the
geometry of the trap. The non-quadratic terms are expected to be small
and can be dealt with perturbatively. The energy and shape of the
condensate change when one includes these terms. It has been shown
\cite{sam} that in terms of the quasiparticle operator $\beta_i$ the
effective Hamiltonian can be written as
\begin{widetext}
\begin{equation}
  H'= const + \sum_{i \neq 0} (\epsilon_i+\Delta \epsilon_i)\beta_i^{\dag} \beta_i+\left\{\left[\sum_{ijk \neq 0} \left [ \xi_{ijk}\beta_i \beta_j \beta_k+ \zeta_{ijk} \beta_i^{\dag} \beta_j \beta_k \right] +\sum_{i \neq 0}\eta_i \beta_i\right]+h.c.\right\},
  \label{eq:H3}
\end{equation}
\end{widetext}
where the $const$ term simply shifts the zero of energy and $\Delta
\epsilon_i$ is the energy shift from first order perturbation theory.
The Hamiltonian given by equation (\ref{eq:H3}) contains terms beyond
the Bogoliubov approximation, showing that we are treating {\it
  interacting} quasiparticles.  We are thus taking into account
important processes such as Landau and Beliaev damping. Landau
processes, in which two quasiparticles collide to form a single
quasiparticle, can not occur at zero temperature because there are no
excited quasiparticles.  However, they are dominant at high
temperatures. Expressions for the coefficients of equation
(\ref{eq:H3}) are given in Appendix \ref{sec:appendix_a}. As we are
interested in finite temperature, the $U$ and $V$ matrices are
calculated self-consistently using the Bogoliubov-de Gennes (BdG)
formalism for each temperature.  From the Hamiltonian (\ref{eq:H3}),
the Heisenberg equation of motion for $\beta_p$ is

\begin{widetext}
\begin{equation}
  \label{eq:beta}
  {\rm i} \dot \beta_p = \frac{\eta_p^*}{\hbar}+\omega_p\beta_p+\sum_{j,k \neq 0}\sigma_{jk}\beta_j \beta_k+\sum_{j,k \neq 0} \rho_{jk}\beta_j^\dag \beta_k^\dag+\sum_{j,k \neq 0} \nu_{jk}\beta_k^\dag \beta_j,
\end{equation}
\end{widetext}
where
\begin{eqnarray*}
\omega_p&=&\left(\epsilon_p+\Delta \epsilon_p\right)/\hbar,\\
\sigma_{jk}&=&\zeta_{pjk}/\hbar,\\
\rho_{jk}&=&\left(\xi_{pjk}^*+\xi_{jkp}^*+\xi_{kpj}^*\right)/\hbar,\\
\nu_{jk}&=&\left(\zeta_{jpk}^*+\zeta_{jkp}^*\right)/\hbar.
\end{eqnarray*}

Equation (\ref{eq:beta}) and its complex conjugate are valid for the
general case. However, a dominant process can be defined by choosing
the appropriate geometry for the trap. In this form, the number of
states involved is reduced \cite{merzbacher}. The trap geometry can be
modified by adjusting the frequencies in the radial and axial
directions independently and the selection of a dominant mode in this
way has been demonstrated \cite{hechen}. In particular, the
observation of a Beliaev process has recently been reported for a
scissors mode, where one mode is resonantly coupled to two modes of
half the original frequency \cite{beldamp}; this is the kind of
process that we seek to model in the present paper. We note that terms
in $\xi$ and $\eta$ describe the spontaneous decay of quasiparticles
and thus do not generally conserve energy. Therefore, we will only
keep terms containing $\sigma$ and $\nu$.  More details of these
arguments can be found in reference \cite{martin}, and the main points
are summarised in Appendix \ref{sec:appendix_b}.  For a Beliaev
process, the equations of motion for modes $p=1$ and $p=2$ are

\begin{eqnarray}
  \label{eq:twomodes1}
  \dot \beta_1 & = & -{\rm i}\omega_1\beta_1-{\rm i}\nu_{21}\beta_2\beta_1^{\dag},\\
  \label{eq:twomodes2}\dot \beta_2 & = & -{\rm i}\omega_2\beta_2-{\rm i}\frac{\nu_{21}}{2}\beta_1\beta_1.\\
\end{eqnarray}
plus their complex conjugates.  The factor of 2 that appears in
equation (\ref{eq:twomodes2}) comes from the fact that
$\sigma_{11}=\nu_{21}/2$.

Considering the system at a temperature $T=20$ nK, with $N=10000$
particles and a trap frequency $\omega_{trap}=2\pi\times 100$ Hz with
spherical geometry and using $^{87}$Rb atoms for which the s-wave
scattering length $a=10 a_0$, where $a_0$ is the Bohr radius, the
coefficient $\nu_{21}$ equals $1.90\times 10^{-2}$ in trap units.  The
equations of motion for $\beta_1$ and $\beta_1^{\dag}$ are profoundly
affected by temperature through the values of $\nu_{21}$.

Equations of this sort have extensively been studied in quantum
optics. In this case we apply them in order to study squeezing in the
quasiparticle excitations and its temperature dependence. It is
important to mention that the damping of excitations has earlier been
described in terms of nonlinear mixing \cite{NLM}; this work differs
from that description as the operator nature of the quasiparticle
annihilation operator $\beta_i$ is retained, and the calculation is
not restricted to the quadratic approximation. We are in effect
extending the work of Ref. \cite{NLM}, such that spontaneous quantum
processes are included.

\section{Temperature dependent coupling process}
\label{sec:pdc}
\subsection{Non-depleted regime}
\label{sec:nondeplete}

In the case in which mode 2 has a much larger population than mode 1,
we may approximate the operator $\beta_2$ by a c-number $b_2$ and
ignore the depletion of mode 2. The solutions of the Heisenberg
equations of motion for the operators $\beta_1$ and $\beta_1^\dag$ are
then given by

\begin{equation}
  \label{eq:solqp}
  \left( \begin{array}{c}
    \beta_1(t)\\ \beta_1^\dag(t)
  \end{array} \right) = \left(
  \begin{array}{rr}\cosh(\Omega t) & -{\rm i} \sinh(\Omega t) \\ {\rm i}\sinh(\Omega t) & \cosh(\Omega t) 
  \end{array} \right) \left(
  \begin{array}{c}
    \beta_1(0) \\ \beta_1^{\dag}(0)
  \end{array} \right),
\end{equation}

where $\Omega=\nu_{21}b_2$, and we have chosen the phases such that
the coefficients are real. This represents the physical situation
where the pump is continuously driven by a resonant excitation. It is
important to mention that the approximation is no longer valid when a
noticeable down-conversion has occurred.

Solutions (\ref{eq:solqp}) are equivalent to equations obtained in
the description of degenerate parametric down-conversion in quantum
optics. This process is well-known to be an efficient source of
squeezed states. In this case, the corresponding squeezing parameter is

\begin{equation}
  \label{eq:sqparameter}
  \tau=\Omega t,
\end{equation}
which is temperature dependent via $\Omega$.

Analogously, one can define the quadrature operators for $\beta$ and
$\beta^\dag$ and calculate the variances

\begin{eqnarray}
  \label{eq:x1}
  (\Delta X(t))^2&=&\frac{(2N_1 +1)}{4} \exp({-2\Omega t}),\\ 
  \label{eq:x2}
  (\Delta Y(t))^2&=&\frac{(2N_1 +1)}{4} \exp({2\Omega t}),
\end{eqnarray}  
where $N_1$ is the number of particles in mode 1. In optics the
quadrature operators are well-defined quantities corresponding to the
amplitude and the phase of the electromagnetic (EM) oscillation. In a
similar fashion, we give our quadrature operators $X$ and $Y$ the
interpretations of amplitude and phase of oscillations. The dependence
of equations (\ref{eq:x1}) and (\ref{eq:x2}) implies that the amount
of squeezing is related to the number of lower-mode atoms present.
Consequently, it is connected to the temperature of the ultracold
atoms in the condensate \cite{rich}, defined in terms of the initial
Bose-Einstein (BE) distribution of quasiparticles.

However, in an experiment what one measures some correlation
functions, for instance $\langle \beta_1^\dag\beta_1 \rangle$.
Assuming an initial number state in the quasiparticle basis, some
important correlation functions are given by:

\begin{eqnarray}
  \label{eq:corr}
  \langle \beta_1(t) \beta_1(t) \rangle &=& -{\rm i}\left(N_1+\frac{1}{2}\right)\sinh(2\Omega t),\\
  \langle \beta_1^\dag(t) \beta_1(t) \rangle &=& \left(N_1+\frac{1}{2}\right)\cosh(2\Omega t)-\frac{1}{2}.\label{eq:N1}
\end{eqnarray}

A plot of the correlation function (\ref{eq:corr}) is shown in Figure
\ref{fig:betasb} a). A non-zero value for this correlation function
implies the presence of squeezing and hence of entanglement between
the two modes. The quantity given by equation (\ref{eq:N1}) carries
information about the behaviour of the population in mode 1 and it is
directly measured as the amplitude of oscillation. A plot of this
correlation function as a function of time $(t)$ and temperature $(T)$
is given in Figure \ref{fig:betasb} b). At $t=0$ the behaviour is
determined by the BE distribution of $N_1$. As time evolves the
population of the lower mode increases.

\subsection{Depleted regime}
\label{sec:depleted}

Let us now include the effect that depletion of mode 2 has in the
description of the coupling process. In order to tackle the problem,
we will consider the equations of motion for the number operators
$N_i=\beta_i^{\dag}\beta_i$.  They can be calculated through the
Heisenberg equations (\ref{eq:twomodes1}-\ref{eq:twomodes2}), namely
\begin{eqnarray}
  \label{eq:N_1} 
  \frac{{\rm d}{N_1}}{{\rm d} t}&=&{\rm i}\nu_{21}\left(\beta_2^\dag\beta_1\beta_1-\beta_1^\dag\beta_1^\dag\beta_2\right),\\ 
  \label{eq:N_2}
  \frac{{\rm d}{N_2}}{{\rm d} t}&=&{\rm i}\frac{\nu_{21}}{2}\left(\beta_2\beta_1^\dag\beta_1^\dag-\beta_2^\dag\beta_1\beta_1\right).
\end{eqnarray}

It is possible to uncouple the equations (\ref{eq:N_1}) and
(\ref{eq:N_2}) using the constant of motion $A=N_1+2N_2$ and
calculating the second derivative. The uncoupled equations are given
by
\begin{eqnarray}
   \label{eq:uncoupN1}
   \frac{{\rm d}^2{N_1}}{{\rm d} t^2}&=&\nu_{21}^2(-3N_1^2+2AN_1+A),\\
   \label{eq:uncoupN2}
   \frac{{\rm d}^2{N_2}}{{\rm d} t^2}&=&\frac{\nu_{21}^2}{2}(12N_2^2-8AN_2+A^2-A).
\end{eqnarray}
When the population of mode 2 is large, it is possible to determine
the time evolution by taking the average of equation
(\ref{eq:uncoupN2}). In that case we have a c-number second-order
differential equation with the following initial conditions
\begin{eqnarray}
  \label{eq:dotN2} 
  \left.\frac{{\rm d}N_2}{{\rm d}t}\right|_{t=0}&=&0,\\ 
  \label{eq:N20} 
  N_2(0)&=&N_{20}; 
\end{eqnarray}

The solution can be expressed as follows
\begin{widetext}
\begin{equation}
\label{eq:sol1N2} 
N_2(t) = \left\{ 
\begin{array}{ccc} 
    N_{20}+(\alpha_2-N_{20}){\rm sn}^2\left(\frac{\nu_{21}}{2}\sqrt{\frac{\alpha_1-N_{20}}{6}}t,\sqrt{\frac{\alpha_2-N_{20}}{\alpha_1-N_{20}}}\right), \text{\ \  for } N_{20}<\alpha_2,\\ 
\\ 
    \alpha_1+(N_{20}-\alpha_1){\rm nd}^2\left(\frac{\nu_{21}}{2}\sqrt{\frac{\alpha_1-\alpha_2}{6}}t,\sqrt{\frac{N_{20}-\alpha_2}{\alpha_1-\alpha_2}}\right), \text{\ \  for } N_{20}>\alpha_2,
\end{array}
\right. 
\end{equation} 
\end{widetext}
where ${\rm sn}(u,k)$ and ${\rm nd}(u,k)$ are the standard Jacobi
elliptic functions \cite{byrd}, and the coefficient $\alpha_1$ and
$\alpha_2$ are the roots of the quadratic polynomial
$P(N_2)=-N_2^2+(A-N_{20})N_2-N_{20}^2-(A^2-A)/4+N_{20} A$. It has been
already mentioned that the coefficient $\nu_{21}$ is temperature
dependent; therefore the population will change accordingly. The
initial population was calculated for different temperatures as
prescribed by the BE distribution. An initial driving was also taken
into account. The solution was then calculated using these parameters;
a plot is shown in Figure \ref{fig:population} as a function of
temperature and time.

The connection between the theory and experiment can be accomplished
by observing that the initial part of the population surface can be
fitted by a curve of the form $A_1\cos(\Gamma t)$. The parameter
$\Gamma$ is related to the solution (\ref{eq:sol1N2}) as
\begin{equation}
  \label{eq:gammanu}
  \Gamma=\nu_{21}\sqrt{\frac{(\alpha_1-N_{20})(N_{20}-\alpha_2)}{12 N_{20}}}.
\end{equation}

Following this line of analysis, a fitting routine was used to obtain
a plot of $\Gamma$ as a function of temperature (Figure
\ref{fig:gamma}). We can see from the plot that for lower temperatures
the influence of the initial population is preeminent. For higher
temperatures the coupling coefficient $\nu_{21}$ dwindles quickly,
thus the effect of the initial population is hidden. It can clearly be
seen that there is a competition between the initial population and
the coupling coefficient. The quantity $\Gamma$ is notable in that it
is a readily measurable temperature dependent quantity that links to the
squeezing of quasiparticles.

\section{Discussion} 
\label{sec:discussion} 

The detection of squeezed states in the case of the electromagnetic
field can be achieved through well-known techniques, e.g. homodyne
detection. On the other hand, scattering experiments have also been
proposed as a way to detect squeezing in bosons \cite{yurke}. The
proposal makes use of a particle scattering off a boson field.
Provided the particle is prepared in the proper input state, it can
either absorb or emit a boson, and these two scattering processes can
coherently interfere when the boson field is in a squeezed state. The
change in the rate of scattering into a particular state gives a
signature that the field has been squeezed. A schematic setup for the
detection of squeezing is shown in Figure \ref{fig:detect}: An
incoming beam of frequency $\omega_{-}$ is divided into two by a beam
splitter, with one of the beams frequency shifted to $\omega_{+}$,
which may be achieved through acusto-optic modulation. The two beams
are made incident into the boson field, e.g.  BEC. Particles scattered
off the squeezed field are detected by the counter. This would provide
an experimentally feasible way to directly measure squeezing in BEC.


In conclusion, we have described a process analogous to parametric
down-conversion in trapped Bose-Einstein condensates. A fully second
quantised order theory has been used to depict the condensate at
finite temperature. The process corresponds to Beliaev damping, where
a quantum of higher energy is divided into two quanta of lower energy.
The coupling process is temperature dependent; this implies that any
amount of squeezing in the elementary excitations of the condensate
would give a direct indication of the temperature.

\begin{acknowledgments} 
  We would like to thank Dr. M. Rusch for helpful discussions. This
  work has been supported by CONACyT, EPSRC, Royal Commission for the
  Exhibition of 1851 and the EU.
\end{acknowledgments}

\appendix 
\section{} 
\label{sec:appendix_a} 
The coefficients that appear in equation (\ref{eq:H3}) written in the
position representation (using the contact potential) are determined
by
\begin{widetext}
\begin{eqnarray}
  \xi_{ijk} &=& \sqrt{N_0}U_0 \int {\rm d}^3{\bf r} \left[\psi_0^*({\bf r}) v_{i}({\bf r}) u_{j}({\bf r}) u_{k}({\bf r})+\psi_0({\bf r}) u_{i}({\bf r}) v_{j}({\bf r}) v_{k}({\bf r})\right],\\
  \zeta_{ijk} &=& \sqrt{N_0}U_0 \int {\rm d}^3{\bf r} \left\{\psi_0^*({\bf r})\left[u_{i}^*({\bf r}) u_{j}({\bf r}) u_{k}({\bf r})+v_{i}^*({\bf r}) v_{j}({\bf r}) u_{k}({\bf r})+v_{i}^*({\bf r}) u_{j}({\bf r}) v_{k}({\bf r})\right]\right.\nonumber\\
  & & +\left.\psi_0\left[u_{i}^*({\bf r}) u_{j}({\bf r}) v_{k}({\bf r})+u_{i}^*({\bf r}) v_{j({\bf r})} u_{k}({\bf r})+v_{i}^*({\bf r}) v_{j}({\bf r}) v_{k}({\bf r}) \right ]\right\},\\
\eta_i &=& -\sum_{q \neq 0}\left(\zeta_{qqi}+\zeta_{qiq}\right)N_q.
\end{eqnarray}
\end{widetext}
where the indices $i,j$ and $k$ are labels that denote eigen-energy
levels. For a 3-D condensate, the level $i$ is an implicit notation
for the quantum numbers $n,l$ and $m$.

\section{}
\label{sec:appendix_b}
To prove that Beliaev damping is indeed the dominant process with the
given trap geometry discussed in the paper, it is only necessary to
calculate various contributions from the framework of second order
perturbation theory. The energy shift due to various higher order
terms is given by the expression
\begin{subequations}
\begin{eqnarray} 
\Delta E(p)&=&-\sum_{ij \neq 0}\frac{|\xi_{pij}^{P3}|^2}{2(\epsilon_p+\epsilon_i+\epsilon_j)}[1+N_i+N_j]\label{eq:A}\\ 
&&+ \sum_{ij \neq 0}\frac{|\zeta_{pij}+\zeta_{pji}|^2}{2(\epsilon_p-\epsilon_i-\epsilon_j)}[1+N_i+N_j]\label{eq:B1}\\
&&+ \sum_{ij \neq 0}\frac{|\zeta_{ijp}+\zeta_{ipj}|^2}{\epsilon_p-\epsilon_i+\epsilon_j}[N_j-N_i],\label{eq:B2}
\end{eqnarray} 
\end{subequations} 
where line (\ref{eq:A}) indicates spontaneous decay of quasiparticles.
Here the coefficient $\xi_{pij}^{P3}$ is defined as a sum over
permutations of the three indices in $\xi_{pjk}$, i.e.
$\xi_{pijk}^{P3}=\xi_{pjk}+\xi_{pkj}+\xi_{jpk}+\xi_{jkp}+\xi_{kpj}+\xi_{kjp}$.
Lines (\ref{eq:B1}) and (\ref{eq:B2}) give Beliaev and Landau damping
processes respectively. Typical values are given in Ref. \cite{martin}
and owing to the resonance in the denominator, (\ref{eq:B1}) may be
made dominant with a correct choice of experimental parameters
\cite{hechen}.

\begin{center}
  \newpage {\Large Figure Captions}
\end{center}
\begin{enumerate}
\item \label{fig:diag} Schematic diagram of energy levels for a BEC in
  a trap.  In the Beliaev process, a quasiparticle of frequency
  $\omega_2$ interacts with the ground state generating two
  quasiparticles of frequency $\omega_1$ that divide the initial
  energy equally. This process is analogous to the optical parametric
  down-conversion.
  
\item \label{fig:betasb} a) Surface plot of the average $\langle
  \beta_1\beta_1 \rangle$. A non-zero value of this quantity implies
  the presence of squeezing and hence entanglement. b) The average
  $\langle \beta_1^\dag\beta_1 \rangle$ describes the behaviour of the
  population in mode 1. This solution is valid within the non-depleted
  regime. As discussed before, the behaviour at $t=0$ is determined by
  the BE distribution function.
   
\item \label{fig:population} The population $N_2$ as a function of
  temperature and time.
  
\item \label{fig:gamma} Plot of $\Gamma$ against temperature. This
  parameter is obtained by fitting a curve of the form $A_1
  \cos(\Gamma t)$ corresponds to the initial part of the envelope of
  the population $N_2$ shown in figure \ref{fig:population}.
  
\item \label{fig:detect} Schematic figure for the detection of
  squeezing by particle scattering.
\end{enumerate}

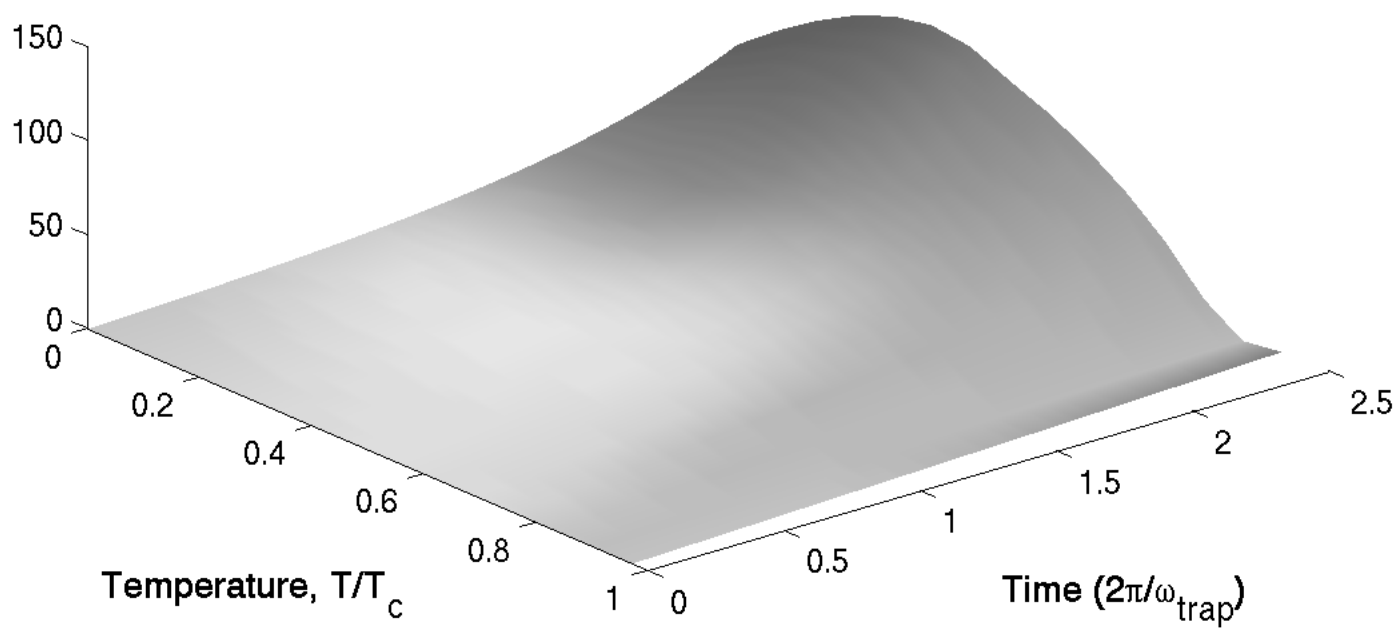
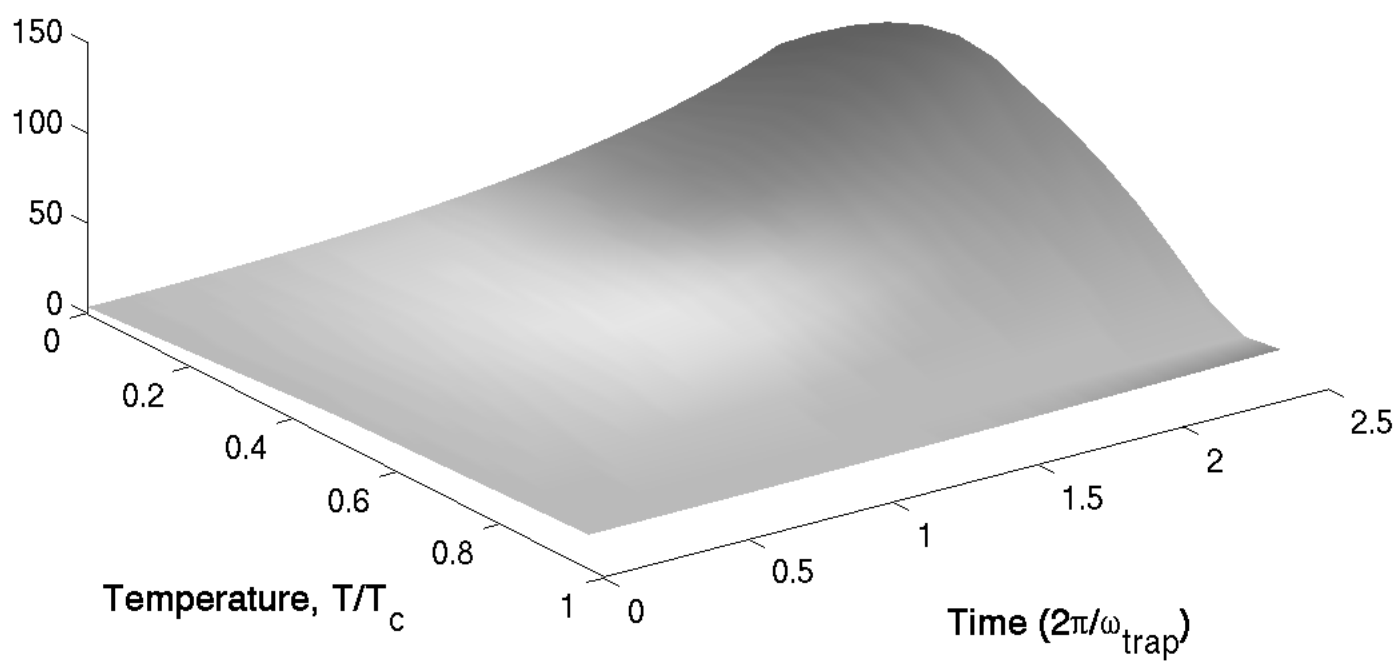
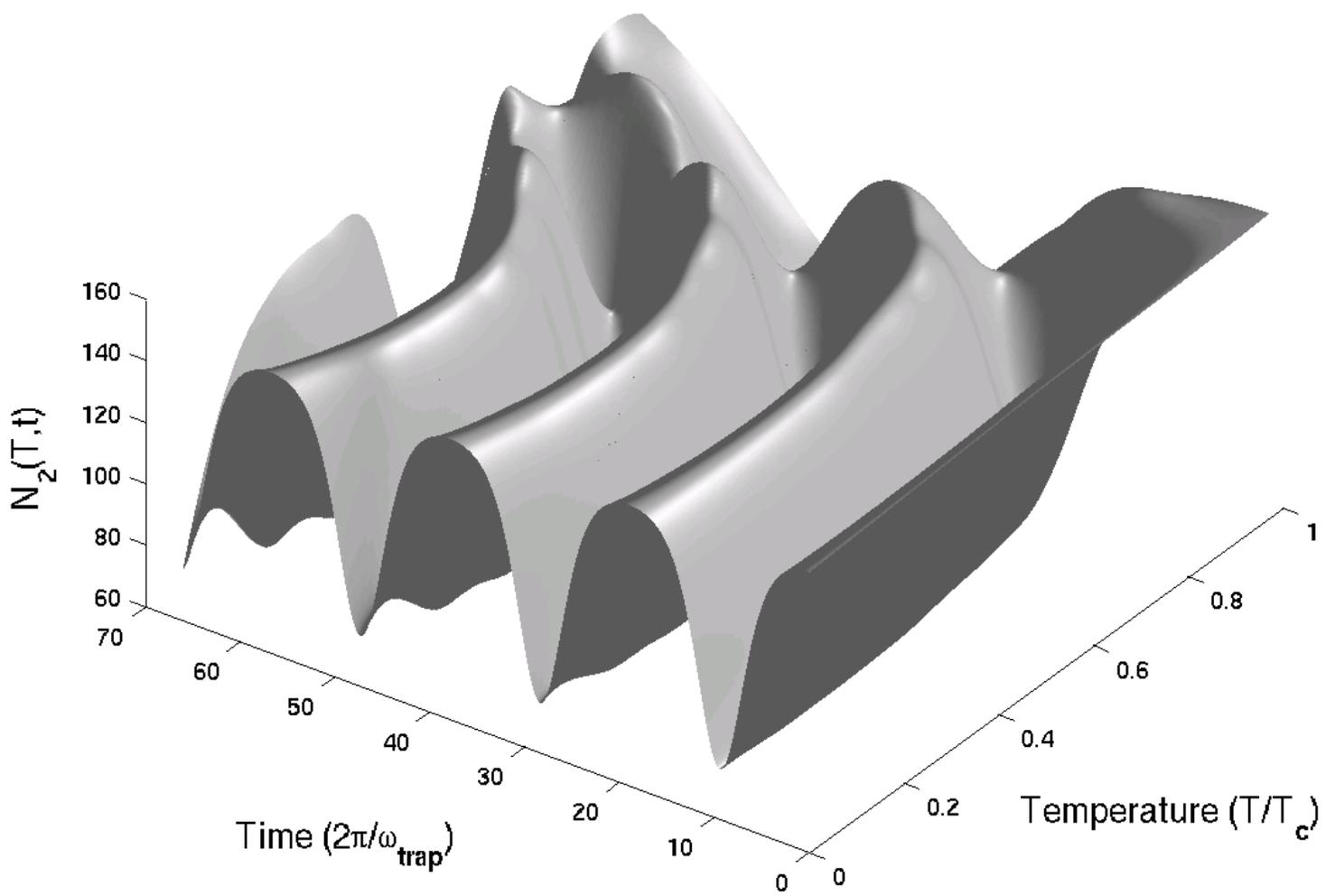
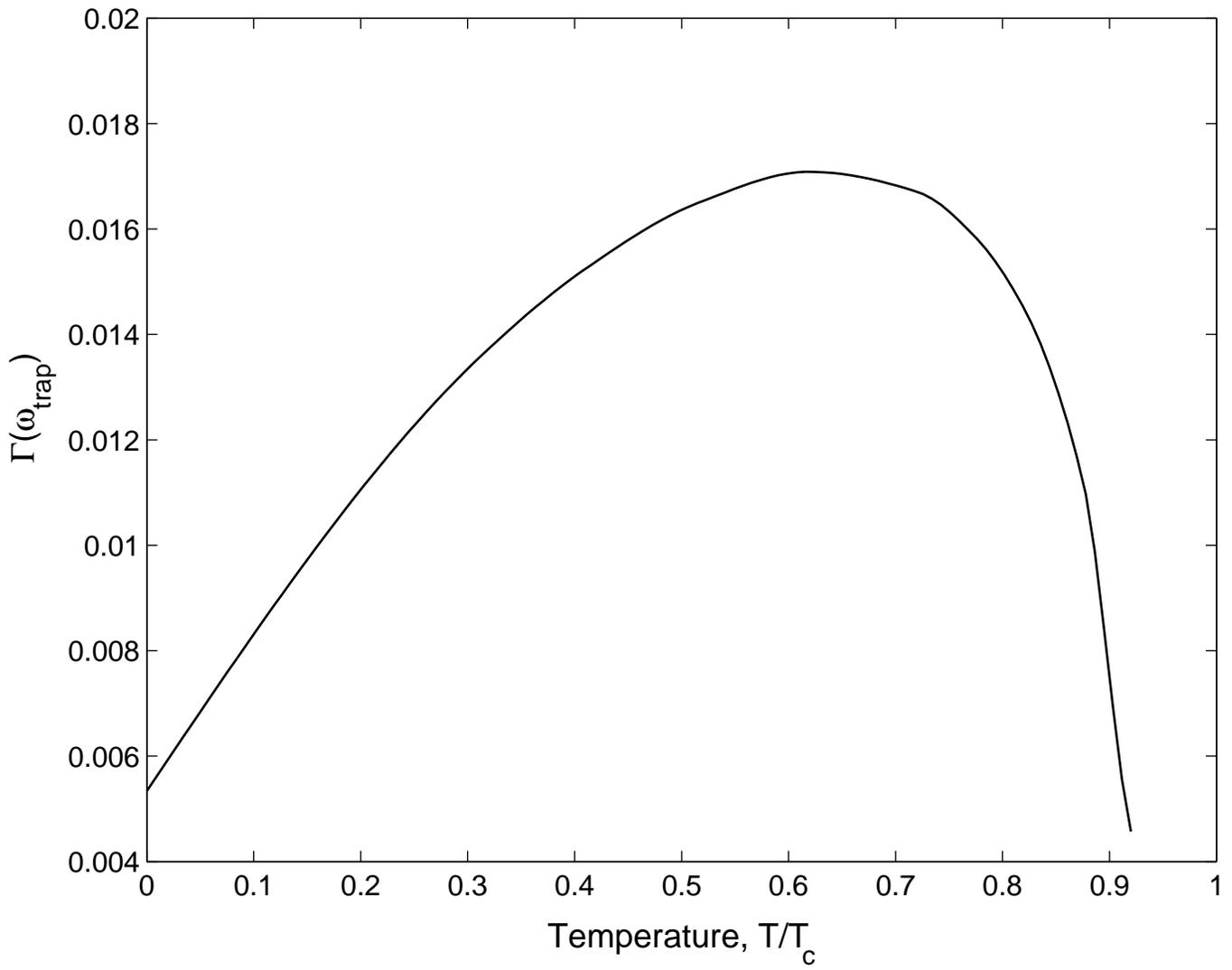
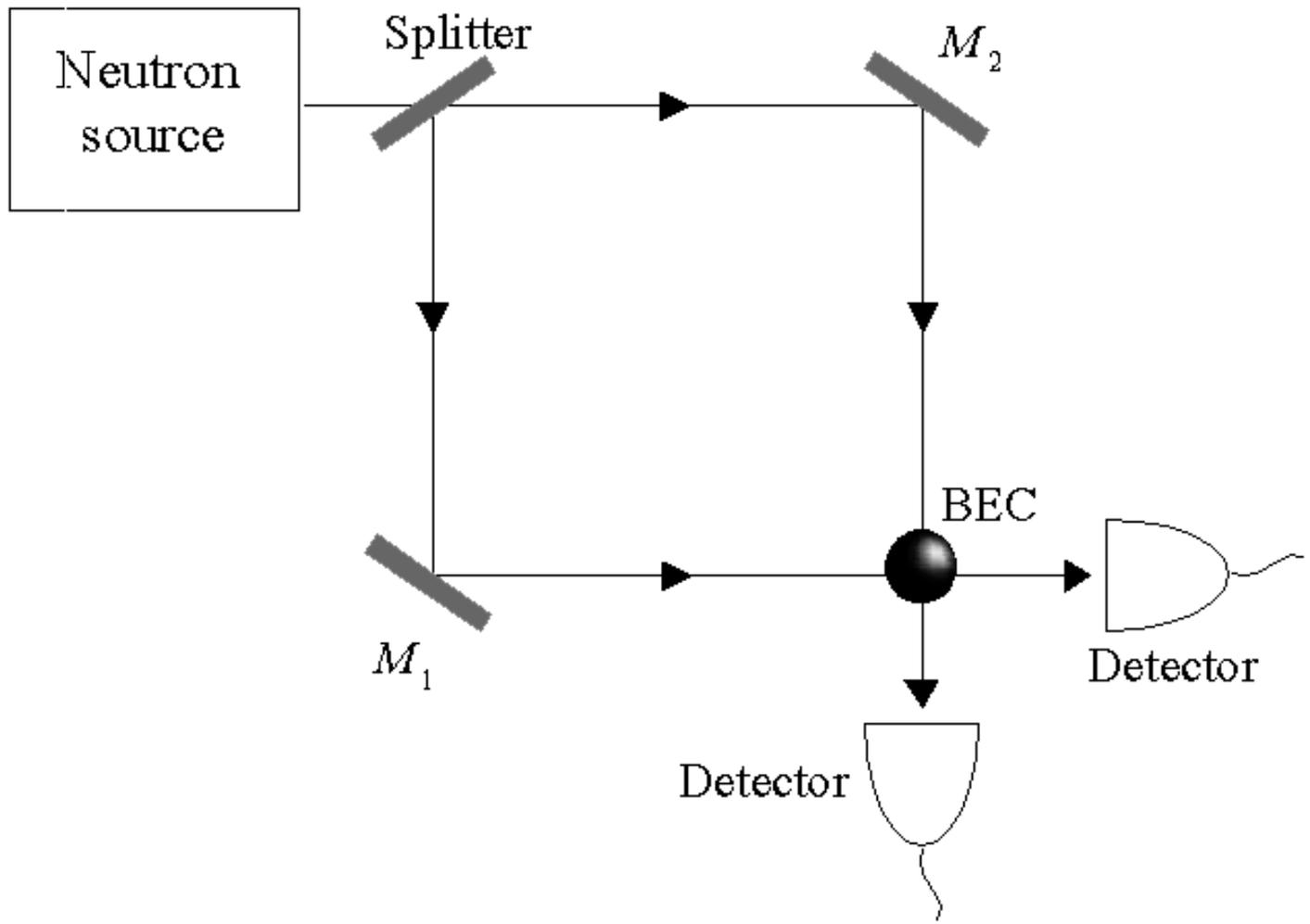
\end{document}